# Dual-Drive Directional Couplers for Programmable Integrated Photonics

DANIEL PÉREZ-LÓPEZ,* ANA M. GUTIERREZ, ERICA SÁNCHEZ,
PROMETHEUS DASMAHAPATRA, AND JOSÉ CAPMANY

*Photonics Research Labs, iTEAM Research Institute, Universitat Politècnica de València, Spain.*
*dperez@iteam.upv.es*

**Abstract:** A novel class of photonic integrated circuits employs large-scale integration of combined beam splitters and waveguides loaded with phase actuators to provide complex linear processing functionalities that can be reconfigured dynamically. Here, we propose and experimentally demonstrate a thermally-actuated Dual-Drive Directional Coupler (DD-DC) design, integrated in a silicon nitride platform, functioning both as a standalone optical component providing arbitrary optical beam splitting and common phase as well as for its use in waveguide mesh arrangements. We analyze the experimental demonstration of the first integration of a triangular waveguide mesh arrangement, and the first DD-DC based arrangement along with an extended analysis of its performance and scalability.

## 1. Introduction

Microelectronics has become one of the pillars of digital economics in the early 21$^{st}$ century. Current and emerging applications demand information processing at a faster speed and bandwidth, exposing a potential physical limitation of electronic systems. The cooperative use of electronics and photonics is being studied and applied as an appealing solution to overcome future performance limits, leveraging on the best of two worlds for both digital and analog processing.

Integrated photonics is the science and technology that enables the integration of a large number of waveguide elements and specific devices or performance blocks in order to enable optical signal processing on a chip. Traditionally, both industry and academia have mainly focused on the design and optimization of Application Specific Photonic Integrated Circuits (ASPICs) whereby all the stages involved in the development of a PIC are tailored to optimize the chip performance, power budget, electrical power consumption, and footprint [1]. This strategy involves the optimization of photonic-based systems through multiple time-consuming cycles of custom design, fabrication, packaging and testing, leading to solutions that, in most cases, are far from being cost-effective for low and moderate volumes. Only very large volumes benefit from economies of scale, but as of yet, such applications are limited to data centre interconnects and transceivers [2].

Most of circuit designs for linear processing applications employ combinations of waveguide interconnections, beamsplitters, like directional couplers (DC), and phase actuators. The performance of each DC is highly dependent on its wavelength of operation, the waveguide geometry and the refractive indices of the materials [3]. These aforementioned dependencies, in turn, are strongly influenced by nanometre scale geometrical fluctuations arising from factors such as fabrication deviations, variations in grown layer thickness, etc. which ultimately result in an equivalent shift of the wavelength of operation of such devices. On the other hand, reconfigurable phase shifters are circuit actuators that introduce a local phase shift of the optical signal when a control signal (typically electrical) is applied. These elements exploit different material properties to change the effective index of the waveguide: thermo-optic, electro-optic, magneto-optic, opto-mechanical, stress, etc. The integration of phase actuators is essential for the construction of functioning PICs, as the offered tunability provides for a scope to combat any spurious effects arising from undesired geometric variation obtained in the fabricated circuit. For example, a fixed optical filter can be tuned by adding extra phase shifters at key points in the optical circuit thus, compensating for the fabrication induced phase drifts [4].

In parallel, a paradigm shift in PIC design explores the development of a multifunctional programmable circuit, where a common integrated optical hardware configuration is programmed to implement a variety of functionalities that can be elaborated for basic or complex operations in many application fields [5-8]. In this respect, this scheme might compromise the overall power consumption, power budget and footprint in order to provide a platform with an unprecedented degree of flexibility and versatility, features which are then inherited by the systems in which it is used. This approach enables a new generation of field-programmable PICs that will potentially offer cost-effective and ready-to-use solutions and allow upgradable photonic-based systems that provide post-compensation after a failure event [9]. Most of the experimental demonstrations to date rely a powerful reconfigurable optical core comprising of the interconnection of multiple instances of a tunable coupler in the form of dual-drive Mach-Zehnder Interferometer with a phase shifter in each of its two arms and characterized by its simplicity, robustness, yield and performance [5-8]. However, to ensure the future scalability of these systems, research should be done in the optimization of this Tunable Basic Units (TBU) architectures and tuning mechanisms to reduce their insertion loss, footprint, optical crosstalk, and power consumption. In this paper, we study the Dual-Drive Directional Coupler (DD-DC) as a candidate for application-specific PICs and waveguide mesh arrangements. In Section 2 we introduce the device configuration and operating principle. In Section 3 and 4 we present the experimental demonstration of a standalone component, in the context of conventional PIC layouts and as the key element in a waveguide mesh arrangement, respectively. Finally, in Section 5 we analyze the associated scalability issues and discuss their future evolution and applications.

## 2. Dual-Drive Directional Couplers

The directional coupler is one of the most frequently employed basic building blocks present in any PIC. Usually, it is designed to operate as a beam splitter characterised by a desired fixed optical power splitting ratio $K$ at a certain wavelength. The signal in one waveguide is completely transferred to a parallel waveguide after each periodic length $L_c$ by a factor referred to as coupling constant $\kappa$, which depends on the wavelength of operation, the waveguide geometry and the refractive indices of the materials [3]. These dependencies result in the DCs being susceptible to fabrication errors that can change the designed wavelength of operation, bandwidth and uniformity.

The integration of a phase tuning mechanism in one of the arms enables the tuning of the effective index difference between the two waveguides, and therefore the resulting coupling coefficient. This capability can be used to provide a tunable splitting coefficient to the designed circuit as well as to compensate for fabrication variations. Standalone Tunable Directional Couplers (TDC) have been demonstrated in polymer materials [10-13], photonic crystals [14] and in Silicon on Insulator (SOI) [15], providing a reconfigurable splitting ratio by enabling a propagation constant difference in the pair of waveguides. The phase-change effect employed has been performed by means of a thermal-tuner placed on top of one of the parallel waveguides or by applying an electro-optic effect to introduce a propagation constant difference between the waveguides. However, the setting of the coupling coefficient introduces an accumulated phase shift at each output [15].

In order to increase the capabilities of TDCs, the integration of a second phase shifter in the other waveguide provides a symmetric design in terms of the possible loss in each arm, as well as both, an independent beam splitting capability and an additional phase shifting capability by inducing a differential and common phase shift, respectively, at each arm [16]. This additional feature can find applications in conventional PICs as well as enable their use as TBUs in waveguide mesh arrangements. A cross-section and top-view of a thermally-tuned Dual-Drive TDC is illustrated in Fig. 1. The analytical model is described in [9].

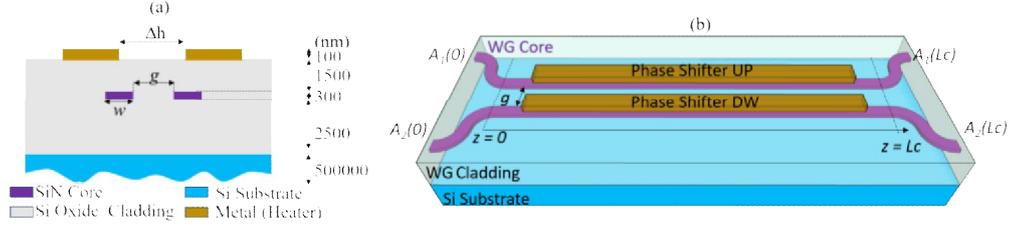

Fig. 1. Dual-Drive Directional Coupler (a) crossection and (b) topview.

## 3. DD-DC Design and Experimental Demonstration

For the characterization of the DD-DC, we designed a chip layout based on a deep-etched waveguide cross section of 1 μm width (*w*) and 300nm height to satisfy the single-mode propagation condition. This geometry has been widely employed in previous Multi-Project Wafer Runs [17]. A mode solver was used, taking in consideration a central wavelength of 1.55 μm and the aforementioned geometric parameters and it yielded a TE mode effective index of 1.5767, group index of 1.9234 and a second-order dispersion of -1.2027 ps/(nm·m).

The chip layout incorporates three layers: the optical waveguide, the metal layer and the thermal/optical isolation trench layer. It is to be noted that the isolation layer involves etching up to the bottom cladding layer, i.e. up to a depth of 1800 μm from the top cladding. The design incorporates the presence of DD-DCs as standalone couplers and tunable couplers in optical ring resonators (ORR) and in a triangular waveguide mesh arrangement of five TBU [5]. One key parameter in the configuration of a DD-DC is the difference between the phases of each waveguide. As mentioned, the propagation conditions are locally modified by a thermal-tuning effect produced by a microheater over the waveguide. The phase shift in a standalone waveguide is proportional to:

$$\Delta\phi = \Delta\beta \cdot L = \frac{2\pi}{\lambda}\Delta n_{eff} L = \frac{2\pi}{\lambda}\frac{dn_{eff}}{dT}\Delta T_{eff} \cdot L \qquad (1)$$

where $\Delta\beta$ is the change in the propagation coefficient, $L$ is the effective length of the phase actuator and $\Delta n_{eff}$ is the change in the effective index, given by the multiplication of the thermo-optic coefficient ($dn_{eff}/dT$) and the effective temperature gradient ($\Delta T_{eff}$) at the waveguide layer. Since the thermo-optic coefficient is one order of magnitude lower than of SOI waveguide, a thermal actuator in silicon nitride requires a $\Delta T \cdot L$ product one order of magnitude greater than that in SOI. This effect translates into the need for considerably longer phase shifters.

Even considering that only one phase shifter is actuated, some tuning effects can produce a strong spread effect which can lead to an induced and unwanted effective refractive index change in the neighbouring waveguide. In the case of thermal actuators, the heat can flow to the adjacent waveguide and modify its propagation conditions. This effect can be modelled by a thermal crosstalk coefficient (CT). Then, the difference between the phase alteration in an actuated waveguide with phase, $\Delta\Phi_1$, and a passive waveguide with phase, $\Delta\Phi_2$, can be expressed as:

$$\Delta\phi_1 - \Delta\phi_2 = (\Delta\beta_1 - \Delta\beta_2) \cdot L = \frac{2\pi}{\lambda}L(\Delta n_{eff,1} - \Delta n_{eff,2}) = \frac{2\pi}{\lambda}\frac{dn_{eff}}{dT}L(\Delta T_{eff,1} - \Delta T_{eff,2})$$
$$= \frac{2\pi}{\lambda}\frac{dn_{eff}}{dT}L \cdot \Delta T_{eff,1}(1 - CT). \qquad (2)$$

where, $CT$ is in the range of 0 and 1. In this case, the proximity of the waveguides means that the value of $CT$ is very large. Hence, in order to obtain a large phase difference between the waveguides, an even larger $\Delta T \cdot L$ product is required. Increasing $\Delta T$ is limited by the properties of the metal layer used to fabricate the micro-heaters over the waveguides. This comes from the fact that increasing the electrical power fed to the resistive heater-element can lead to an irreversible change in the material and even result in its melting and breakdown [18]. In order to ensure that the phase shifters are long enough, we designed passive cross-state configuration DD-DCs with a length of $L_{co}$ = 1278.73 μm, and waveguide gap of 1.7 μm, as obtained from the passive mode-solver simulation displayed in Table I. Some of the designs include variations of $L_{co}$ to compensate for fabrication variations.

TABLE I
SIMULATED DC COUPLING LENGTH VALUES FOR DIFFERENT GAPS AT A WAVELENGTH OF 1.55 μm.

| Gap (μm) | 0.6 | 0.8 | 1.0 | 1.2 | 1.4 | 1.6 | **1.7** |
|---|---|---|---|---|---|---|---|
| (w=1μm), $L_{co}$ (μm) | 58.2 | 101.7 | 180.6 | 311.2 | 543.7 | 945.4 | **1278.73** |
| (w=1.2 μm), $L_{co}$ (μm) | 90.6 | 163.6 | 301.2 | 565.0 | 1002.4 | 1852.3 | TBD |

In the following sections we analyze the measured results for each DD-DC application.

*3.1 DD-DC Standalone configuration*

The first group of three components are $TDC_1$, $TDC_2$ and $TDC_3$. They are standalone DD-DCs with all ports accessible via edge-couplers. In order to compensate from fabrication variations and test the design tolerances, we have parametrized each coupler length to be: $TDC_1$: $L_{co}$ -30 μm, $TDC_2$: $L_{co}$, $TDC_3$: $L_{co}$ +30 μm, all of them having separations of 1.7 μm between waveguides and 2.5 μm between micro-heaters. Using this combination of gap and coupling length, theoretically, we achieve a 1% uniformity of the coupling coefficient, across 18 nm.

Fig. 2 shows the relevant section of the fabricate PIC and a labelled illustration of the mask layout. The upper (UP) and lower (DW) heaters and ports are labelled correspondingly.

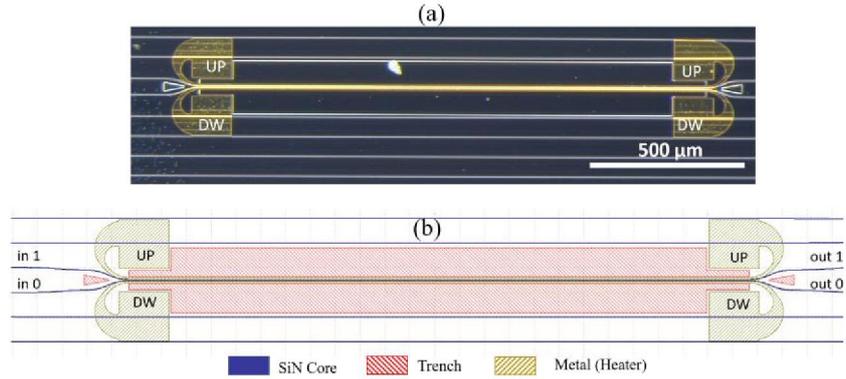

Fig. 2. Picture and labelled mask layout of standalone DD-TDC

In the design, we introduced the use of isolation trenches to focus the heat flow onto the waveguide and thus, to increase the actuator efficiency. Note that the isolation trenches are separated by more than 5 μm from the optical waveguides in order to prevent fabrication defects on the region close to the waveguide core and to avoid a mode interaction of the trench with the effective index. This safety distance was estimated to be around 3 μm by means of finding the negligible interaction between the mode and the trench through a mode-solver analysis.

Once fabricated, we characterized the components with both a conventional scheme involving a tunable laser and a synchronized optical spectrum analyser (OSA) as well as with a broadband amplified spontaneous emission (ASE) light source. The optical coupling was implemented using both lensed objectives and lensed fibers. In both schemes, we measured edge coupling losses to be within a range of 3 dB/facet with a variability of ±1 dB. In regard to the electrical driving of the thermal actuators we probed the DC pads of each heater and employed a high precision Keithley current source. Until and unless specified, the measurements are single drive, i.e. only one heater is activated at a time.

In Fig. 3a, we illustrate the cross-port spectral response of TDC3, normalized with respect to the response of a 11mm long straight waveguide. We can see that a flat uniformity is maintained over the 6 nm range. Each trace corresponds to a different electrical driving condition. We can see that the power coupled to the opposite waveguide in a passive state is around 3 dB less than the maximum power, meaning that the targeted passive cross-state at the design stage is not achieved. Next, if we set an electrical current in the upper phase shifter, we find a maximum of the total optical power in the cross-port (upper-trace). If we drive the lower heater (DW), we find that the coupling ratio goes in the opposite direction till it reaches a minimum optical coupling at around -30dB. A 0.2-dB ripple is observed in some of the measurements and is associated to the reflections in the measurement setup and the chip facets.

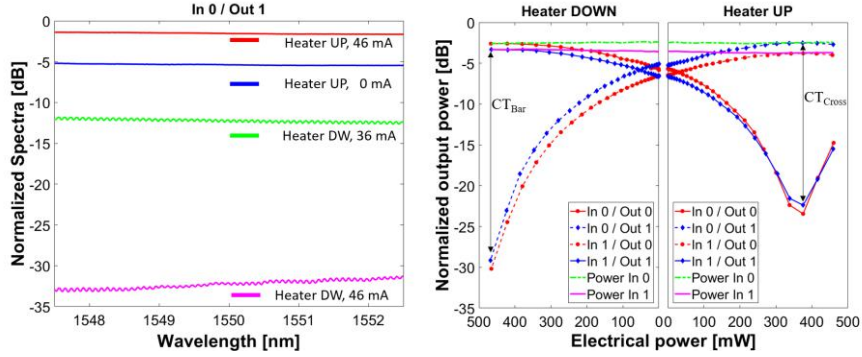

Fig. 3. (a) Spectral response of the DD-TDC 3 and (b) Experimental spectral response of a DD-TDC when upper or lower phase shifters are actuated.

In order to characterize the dynamic behaviour of the structure, we performed continuous sweeps of electrical current with a step of 2 mA in one direction (heater up) and then in the opposite (heater down). For each case we measured the response for all input-output combinations. The cross-port power is directly proportional to the optical power coupling value ($K$) of the DD-DC. From the measurements illustrated in Fig. 3b, it can be seen that the patter is typical of a cross point interferometer and its complementary response, maintaining the conservation of energy during the configuration. We repeated the test for thirteen different TDCs in four different dies and five wafers and concluded that all the devices, but one maintained the same trends as in Fig. 3b. Due to the setup, facet and alignment variation, it is challenging to estimate the insertion loss of the DD-DC, but we can infer that this is well below 0.3 dB and that the conservation of energy is maintained during the dynamic configuration. When we tune the cross or bar states, we measure a portion of the signal that leaks over the non-desired output, known as the optical crosstalk ($CT_{opt}$). Overall, half of the measurements obtained optical crosstalk in the cross and bar states better than 15 dB. The measurements summary is contained in Table II.

From the measurements, we can see that some of the DD-DC shown a non-ideal behaviour with $CT_{opt}$ below 15 dB in either cross or bar. We can see that this behaviour is present in more than the 50% of the measured TDCs, independently of the coupling length. This results from low fabrication tolerance of the DCs. In addition, all the $CT_{bar}$(dB) could be increased if we do not limit the maximum applicable current to 40 mA. However, there was a high-risk of achieving a breaking point when driving the heater with larger electrical currents.

TABLE II
MEASUREMENTS OF STANDALONE DD-DC SUMMARY: CT: OPTICAL CROSSTALK, PE: ELECTRICAL POWER

| item | Wafer | Die | TDC# | CT bar (dB) | CT cross (dB) | Pe bar (mW) | Pe cross (mW) |
|---|---|---|---|---|---|---|---|
| 1 | 1 | 5 | 1 | 12.732 | 14.184 | 431.92 | 317.35 |
| 2 | 1 | 6 | 3 | 15.339 | 14.509 | 455.42 | 284.12 |
| 3 | 2 | 3 | 3 | 29.034 | 11.534 | 399.12 | 445.65 |
| 4 | 3 | 4 | 3 | 16.588 | 11.993 | 528.70 | 165.57 |
| 5 | 3 | 5 | 1 | 4.216 | 17.937 | 429.97 | 162.45 |
| 6 | 3 | 5 | 2 | 12.790 | 16.626 | 453.17 | 242.75 |
| 7 | 3 | 5 | 3 | 25.698 | 16.196 | 454.90 | 346.85 |
| 8 | 4 | 4 | 2 | 6.193 | 13.982 | 465.85 | 100.28 |

| 9 | 4 | 4 | 3 | 10.647 | 22.509 | 316.97 | 489.17 |
|---|---|---|---|--------|--------|--------|--------|
| 10 | 4 | 5 | 1 | 16.557 | 19.834 | 440.22 | 295.85 |
| 11 | 4 | 5 | 2 | 26.704 | 19.775 | 466.47 | 375.27 |
| 12 | 4 | 5 | 3 | 28.600 | 19.828 | 380.60 | 429.20 |
| 13 | 5 | 4 | 3 | 25.313 | 16.398 | 524.87 | 278.77 |

### 3.2 DD-DC in photonic integrated circuits

The DD-DC can be integrated in a wide variety of PICs requiring both, an arbitrary beam-splitting and a phase adjustment. For example, their integration in an optical ring resonator enables the possibility of modifying the notch position and the extinction ratio independently.

In these designs, we have employed the same waveguide separation of 1.7 μm, distance between heaters of 2.5 μm and a coupling length of $L_{co} + \Delta L$ with length variations (ΔL) equal to: -100, -50, 0, 50 μm. These values have been chosen in order to account for the coupling length variation with changes in the waveguide geometry. For these lengths we created a replica with a larger distance between heaters of 3μm, resulting in a total of 8 optical ring resonators. Coupling length has been estimated to vary between ± 100 μm for waveguides with width variations of ± 20 nm. Figure 4 illustrates a picture of the fabricated device and the designed mask layout.

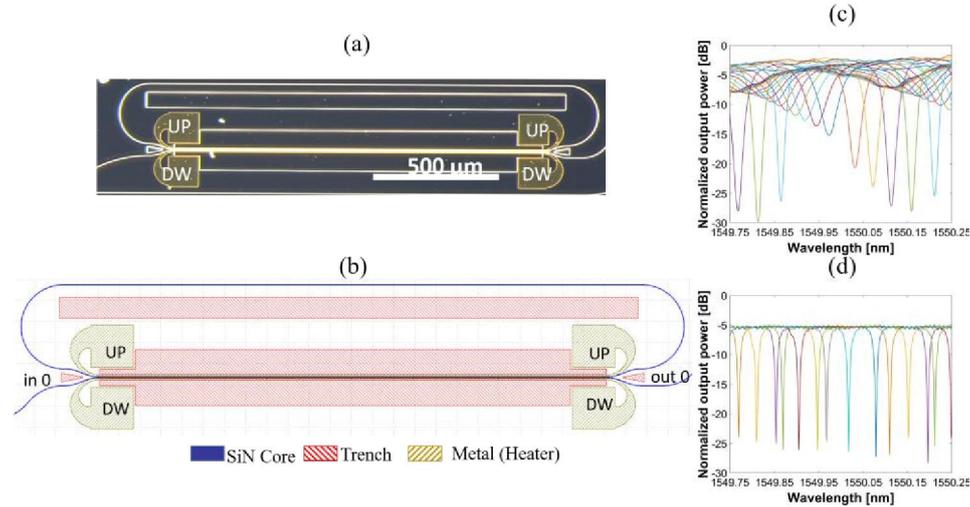

Fig. 4. (a) Picture, (b) labelled mask layout: Ring-loaded version of DD-TDC for its characterization and (c) example of resulting spectral response with single-drive operation, (d) example of notch wavelength tuning with dual-drive operation.

When modifying the coupling coefficient by actuating one of the phase tuners (UP or DW), we produce a power splitting ratio variation as covered in the previous section. This can be employed to modify the extinction ratio of the optical filter. However, as illustrated in Fig. 4c, these changes introduce a common phase shift at the DD-DC outputs, moving the position of the notch. If the splitter is integrated in a PIC, we might want to correct this accumulated phase and set the coupling factor and the accumulated phase independently. Fig. 4d illustrates how a dual-drive operation can be employed to maintain the ER over 19 dB while effectively tuning the wavelength position of the notch by driving the DC in Dual-Drive mode.

Fig. 5 represents the experimental results of a ring resonator based on DD-DCs. Here, we mapped the extinction ratio (ER) and the phase shift versus the electrical power applied to both heaters simultaneously. We can obtain two conclusions from Fig. 5a. First, we see that any arbitrary ER of the ring can be set from 0 to 24 dB by optimally tuning the electrical power of one heater. Secondly, it can be seen that there is a plane where the ER is almost constant once the targeted ER is set and an additional common-drive is applied simultaneously to both heaters. From Fig. 5b, we can extract two complementary conclusions. First, it is confirmed that a single-drive configuration will modify both the ER and the phase shift (notch position). Secondly, the application of the aforementioned common drive is translated to a linear variation of the overall phase shift (notch position).

Experiments, as a whole, determine that once fixed to a targeted ER by driving one of the phase shifters, we can drive both heaters with an additional common power in both actuators to set the notch position without seriously compromising the set ER value.

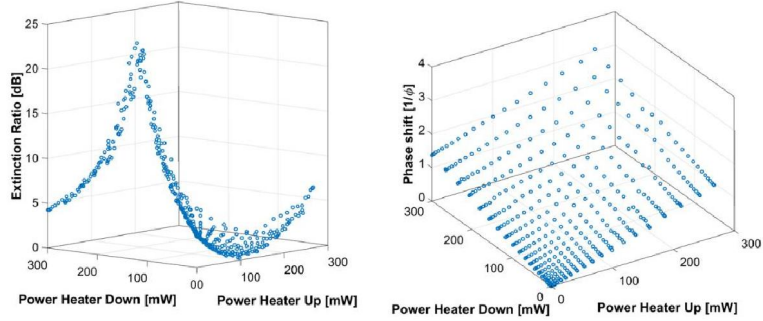

Fig. 5. Extinction Ratio and phase shift variation of a single-notch ORR with a DD-DC (1328.73 µm).

For these measurements, we employed a tunable laser and a synchronized OSA featuring a 1 pm resolution. We measured 10 structures obtaining similar trends and behaviours.

Additionally, we have explored the effects of including a thermal isolation in the form of an air trench between the two adjacent waveguides. For this purpose, we designed an additional set of ORRs with length variations of: 1000, 1200, 1400 and 1600 µm. Figure 6 illustrates the experimental results obtained in a DD-DC with a coupling length of 1200 µm. From the results, we see that this time the ER maximum value is higher, but we did not apply enough power to achieve the bar state that can be characterized by a 0-dB ER value to prevent them from achieving a non-reversible state. In addition, we found that the ER is also limited by the reflections at the facets of the chip couplers.

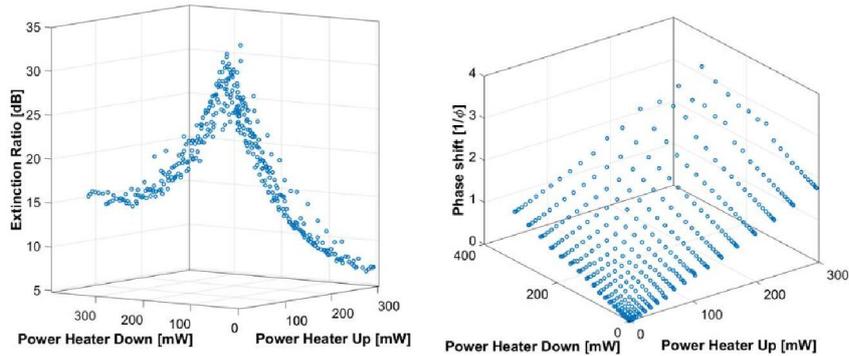

Fig. 6. Extinction Ratio and phase shift variation of a single-notch ORR with a DD-DC (1200 µm).

However, as in the previous section, we found some randomness in the measured performance of both DD-DC designs that might be again related to the fabrication tolerances of the device. As an example, in Fig. 7 we measured the ER while applying single-drive either in the upper or the lower phase shifter. The only difference between both graphs is the distance between metal layers.

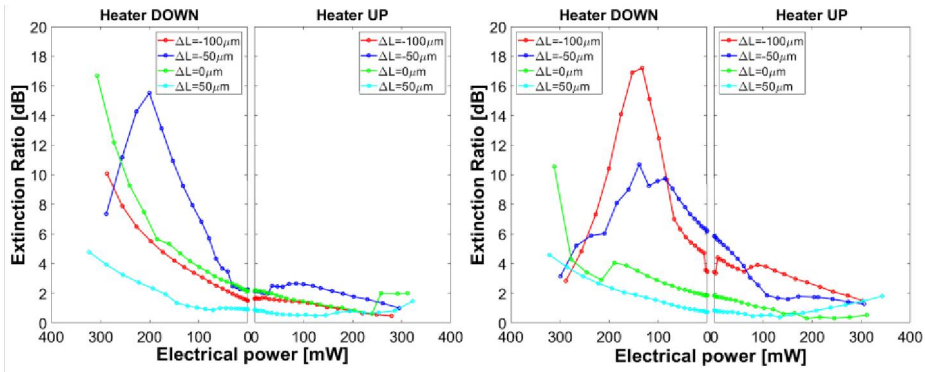

Fig. 7. Extinction Ratio and phase shift variation of a single-notch ORR with a DD-DC for different coupling lengths.

## 4. DD-DCs based Tunable Basic Units for waveguide mesh arrangements

A novel class of programmable photonic ICs employs a massive interconnection of beamsplitters and phase shifter elements to produce multiport interferometric patterns. These arrangements have been proposed based on the replication of Tunable Basic Units (TBU) and their interconnection in a feedforward topology [6] and allowing light recirculation and feedback loops [5,7-9]. Most of the demonstrations rely on TBUs based on thermally-actuated Mach-Zehnder Interferometers loaded with one phase shifter per arm. The performance and versatility of these circuits are limited by the number of integrated TBUs. The scalability of this approach is constrained by the excess loss of each TBU, the optical crosstalk, the power consumption and the footprint. The losses being one of the main limitations, a waveguide mesh circuit becomes impractical when it is made of TBUs with moderate insertion losses (> 0.4 dB) [9] and with optical crosstalk below 20 dB [19]. Due to the considerable number of TBUs that need to be traversed by the signal in a circuit with certain degree of interconnection complexity, a minimal improvement in the TBU insertion loss has a remarkable impact on the overall mesh design and performance. To achieve this, we analyze the possibility of implementing the TBUs using the designed DD-DC instead of a 3-dB MZI.

Figure 8 shows the first implementation of a triangular waveguide mesh arrangement and the first waveguide mesh arrangement exclusively based on DD-DCs. It is characterized by 5 TBUs, building-up two triangular cells, 8 optical ports and 10 phase shifters.

In this simple structure we can program each TBU to build up delay lines (by setting the TBU to cross or bar) and beam splitters of any arbitrary splitting ratio. In addition, a common phase term can be locally added to the outputs of each TBU, as we demonstrated in the previous section. With all these functions, one can discretize conventional PICs into TBU primitives. First, we characterize all the TBU responses by using the mesh characterization protocol recently proposed [20]. Afterwards, we were able to program different structures like delay lines, a single ring resonator, an add-drop filter (transmission and reflection response), and two coupled ring resonators (CROW).

Figure 9a describes the single ring resonator implementation. Here, $TBU_1$ is configured as a tunable coupler, $TBU_2$ and $TBU_3$ are configured in their bar state. The reflection response is obtained employing input port 1 and output port 2 as illustrated in Fig. 9(c1). Furthermore, as illustrated in Fig. 9(b1), we can now exploit the independent phase shifting capabilities of DD-DCs and tune the position of the notch over the full Free Spectral Range without seriously compromising the extinction ratio. Next, we reprogrammed the mesh to perform an add-drop filter by modifying the $TBU_3$ configuration to an arbitrary tunable coupler configuration. As illustrated in Fig. 9c1, we measured the transmission and reflection responses of the ring. Note that we have optimized them separately. The extinction ratios are 24 and 17 dB in the reflection and transmission response, respectively.

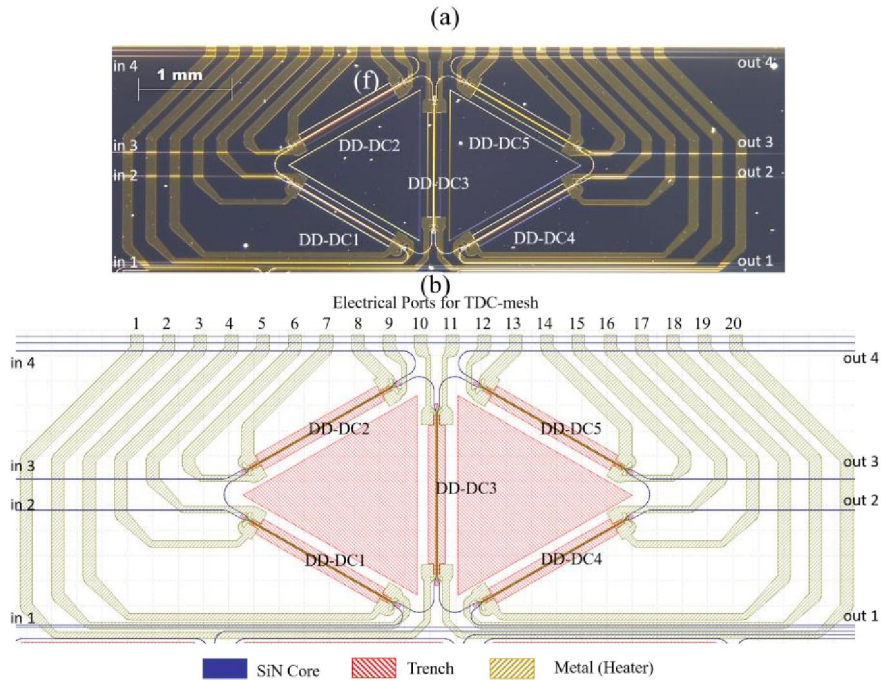

Fig. 8. Picture and labelled mask layout of a triangular waveguide mesh based on DD-DC tunable elements.

The next example illustrates the synthesis of two coupled optical cavities. Again, we employ the input port 1. The reflection response is obtained from output port 2 (Fig. 9(b2)) and the transmission response from output port 7 (Fig. 9(c2)). In this case, the measured ER was 8 and 9 dB, respectively.

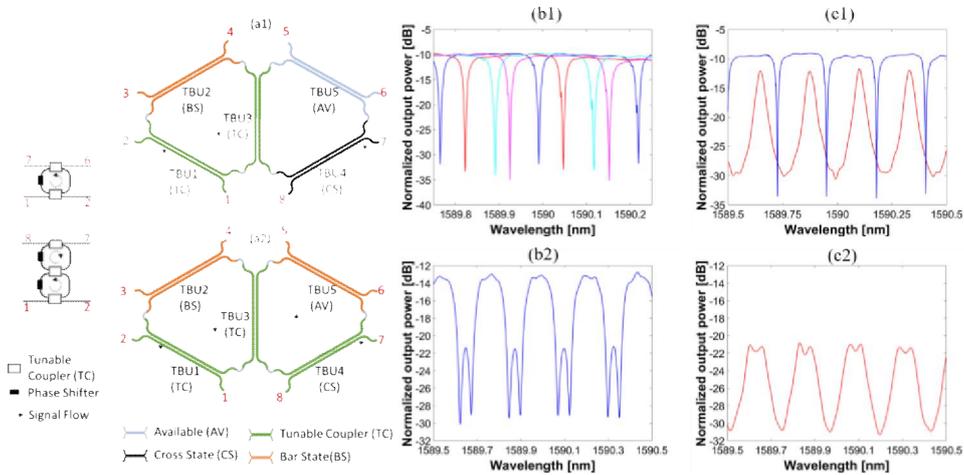

Fig. 9. Triangular feedforward/feed-backward waveguide mesh arrangement based on DD-DC units. (a) Targeted layout and TBU settings, (b) Reflection response of optical circuits, (c) transmission response of optical circuits, First row: single ring resonator and add-drop, Second row: Coupled Ring Resonator Structure.

## 5. Discussion

### 5.1 Theoretical background

The experimental work performed in this paper confirms the behavior of the DD-DC as a 2x2 optical component that enables a tunable splitting ratio and a common phase shift at the outputs. In the experimental results covered in the paper, we measured an unexpected non-symmetrical behaviour between the actuation of the phase shifters. Going deeper into the theoretical framework, directional couplers have been modelled by using coupled-mode equations. These analyses on passive designs have been extended to account for single-drive and for differential modulation schemes and the possibility of altering the propagation conditions of each arm in a differential way [3] or independently [9]. According to these analyses, we would have expected that both phase actuators should provide a symmetric effect when tuning any of the phase shifters in a single-drive mode. However, the proposed theoretical solutions assume that the coupling factor between waveguides ($\kappa$) is kept constant during the tuning and symmetrical (the coupling factor from waveguide 1 to waveguide 2 is equal to the coupling factor from waveguide 2 to waveguide 1). Under these conditions, independently of the phase actuator employed (upper or lower) in a DD-DC, the overall splitting ratio would tend to $\kappa$ equal to zero first and small oscillations would appear around the bar operation point, decoupling the waveguide interaction.

However, the propagation coefficient modification of each waveguide has a direct impact on the coupling factor between the two waveguides ($\kappa_{10}$, $\kappa_{01}$) [21], and might introduce a non-even function for each phase actuator. In addition, with a high probability, the tuning effect will modify the refractive index of the cladding with a different proportion for each waveguide. This is the reason behind the asymmetrical / non-even behaviour of the DD-DC reported in this paper.

*5.2 Large-scale applications*

The proposed architecture can find applications in any PIC requiring reconfigurable beamsplitting and phase shifting. Fig. 10 illustrates examples of their local application in coupled-ring structures, MZIs and waveguide meshes. According to the latter, the proposed architecture potentially reduces the insertion loss of Tunable Basic Units, allowing their scalability and versatility increment. However, future development would require the optimization of the device to increase its tolerance to fabrication variations in order to maintain an optical crosstalk under 20 dB in the targeted wavelength range. In regard to power consumption, the thermo-optic effect requires around 350 mW to tune to either bar or cross state. The high-power requirement arises from the waveguide´s low thermo-optic coefficient and the high tuning crosstalk coefficient in between them. To ensure the future evolution of DD-DC, it should be substituted by alternate tuning mechanisms. The current demonstration is the seed for future works that will potentially combine the proposed architecture with non-volatile phase change materials [22-23], maintaining the decrement in insertion losses and allowing near-zero power consumption. Finally, the footprint of the device is highly influenced by the length required for the desired tuning and should be in the range of hundreds of micrometres to allow high-density integration.

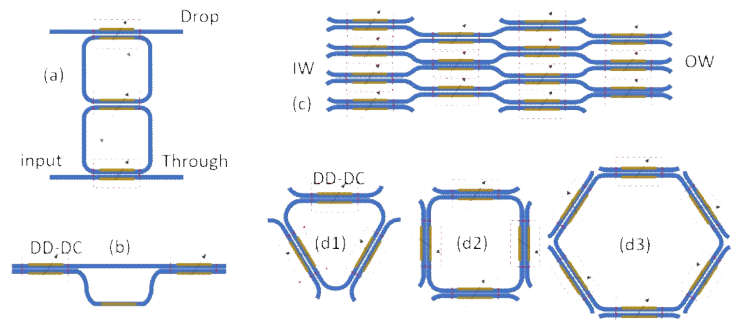

Fig. 10. Photonic integrated circuits with Dual-Drive Directional Couplers. (a) add-drop, (b) Mach-Zehnder Interferometer, (c) arbitrary unitary linear multiport interferometer, (d) feedback-enabled waveguide meshes IW: input waveguides, OW: output waveguides.

## 6. Conclusions

The very-large integration density required in future field-programmable photonic integrated circuits demands the need for optimization of the basic programmable units. Here, we have proposed a new component based on adding two extra phase actuators to the conventional directional coupler. Taking advantage of the thermo-optic effect, we experimentally demonstrate that it is possible to set, independently, the coupling ratio and an overall phase increment that can find applications in programming conventional photonic integrated circuits and in waveguide mesh arrangements. In the case of the latter, they are proposed as a possible option to overcome insertion losses and reduce internal reflections. Future work is required to increase the fabrication tolerances.

## Acknowledgments

The authors acknowledge financial support by the ERC ADG-2016 UMWP-Chip, the Generalitat Valenciana PROMETEO 2017/017 research excellency award, and the COST Action CA16220 EUIMWP.